\documentclass[aps,pra,twocolumn,groupedaddress,floatfix]{revtex4}
\usepackage{graphicx}

\begin{document}

\title{Non-Abelian gauge field theory of the spin-orbit interaction and a perfect spin filter}
\author{Naomichi Hatano}
\affiliation{Institute of Industrial Science, University of Tokyo, Komaba, Meguro, Tokyo 153-8505, Japan}
\email{hatano@iis.u-tokyo.ac.jp}
\author{Ry\=oen Shirasaki}
\affiliation{Department of Physics, Yokohama National University, Tokiwadai, Hodogaya-ku, Yokohama, Kanagawa 240-8501, Japan}
\email{sirasaki@phys.ynu.ac.jp}
\author{Hiroaki Nakamura}
\affiliation{Theory and Computer Simulation Center, National Institute for Fusion Science,
Oroshi-cho, Toki, Gifu 509-5292, Japan}
\email{nakamura@tcsc.nifs.ac.jp}

\date{\today}

\begin{abstract}
We point out that the Rashba and Dresselhaus spin-orbit interactions in two dimensions can be regarded as a Yang-Mills non-Abelian gauge field.
The physical field generated by the gauge field gives the electron wave function a spin-dependent phase which is frequently called the Aharonov-Casher phase.
Applying on an AB ring this non-Abelian field together with the usual vector potential, we can make the interference condition completely destructive for one component of the spin while completely constructive for the other component of the spin over the entire energy range.
This enables us to construct a perfect spin filter.
\end{abstract}



\maketitle

\section{Introduction}
In the present article, we first point out that we can regard the Rashba spin-orbit interaction and the Dresselhaus spin-orbit interaction in two-dimensional semiconductor heterojunctions as a non-Abelian gauge field, or the Yang-Mills field~\cite{Ryder}.
The Yang-Mills field generates a physical field due to which the wave function acquires a spin-dependent phase factor.
This phase is called the Aharonov-Casher phase~\cite{Aharonov84,Mathur92,Oreg92,Balatsky93,Choi93,Oh95,Choi97,Yi97,Yi98,Molnar04,Souma04,Wang05,Konig06} in the context of semiconductor physics and often discussed in connection to the Berry phase~\cite{Berry84,Mathur91,Aronov93,LyandaGeller93,Qian94,Hentschel04}.
We show that the Aharonov-Bohm phase and the Aharonov-Casher phase are straightforwardly understood in terms of the standard non-Abelian gauge field theory in a unified way.

We next show that, applying the non-Abelian gauge field together with the usual vector potential on an mesoscopic interference circuit, we can construct a perfect spin filter from which only one component of the spin comes out.
There have been indeed similar proposals~\cite{Nitta99,Molnar04};
the essential difference here is the explicit specification of the tilt of the spin quantization axis.

The spin-orbit interaction~\cite{Elliott54,Messiah58,Rashba60,Dresselhaus55} in the two-dimensional electron gas attracts much attention recently, particularly because of possibilities of manipulating the spin under an electric field.
There have been many numerical calculations of the conductance in a quantum ring with the Rashba spin-orbit interaction~\cite{Oreg92,Choi97,Yi97,Yi98,Molnar04,Souma04,Wang05,Hentschel04,Frustaglia04}.
However, it is our observation that there are not many guiding principles of designing nanostructures for particular purposes.
We here give a simple way of analyzing the effect of the spin-orbit interaction on the electron wave function, thereby providing a general guiding principle of designing a nano-scale interference circuit.

\section{Non-Abelian Gauge Field Theory}

\subsection{Spin-orbit interaction}
The Hamiltonian with the Rashba spin-orbit interaction is given by~\cite{Engel06}
\begin{equation}
\label{eq010}
\mathcal{H}_\mathrm{RSO}=\frac{1}{2m^\ast}\left({p_x}^2+{p_y}^2\right)
+\alpha\left(p_x\sigma_y-p_y\sigma_x\right)
\end{equation}
and the Hamiltonian with the Dresselhaus spin-orbit interaction is given by
\begin{equation}
\label{eq020}
\mathcal{H}_\mathrm{DSO}=\frac{1}{2m^\ast}\left({p_x}^2+{p_y}^2\right)
+\alpha\left(p_x\sigma_x-p_y\sigma_y\right),
\end{equation}
where $m^\ast$ is the effective mass of the electron and $\alpha$ is the coupling constant of the spin-orbit interaction.
Let us, for the moment, describe our theory in the case of the Rashba interaction;
the discussion is completely parallel for the Dresselhaus interaction.
The key to the analysis below is to transform the Hamiltonian~(\ref{eq010}) in the form
\begin{eqnarray}
\label{eq030}
\mathcal{H}_\mathrm{RSO}=\frac{1}{2m^\ast}\left[
\left(p_x+\theta\frac{\hbar}{2}\sigma_y\right)^2
+\left(p_y-\theta\frac{\hbar}{2}\sigma_x\right)^2\right]
&&
\nonumber\\
+\mathrm{const.},&&
\end{eqnarray}
where
\begin{equation}
\theta\equiv \frac{2m^\ast\alpha}{\hbar}.
\end{equation}
We could regard $\theta$ as a charge and the terms
\begin{equation}
\frac{\hbar}{2}\left(\sigma_y,-\sigma_x\right)
\end{equation}
as the corresponding vector potential, but the essential difference is that the components of the ``vector potential'' do not commute with each other.
Thus we have a non-Abelian gauge field.

\begin{figure}
\centering
\includegraphics[width=0.6\columnwidth]{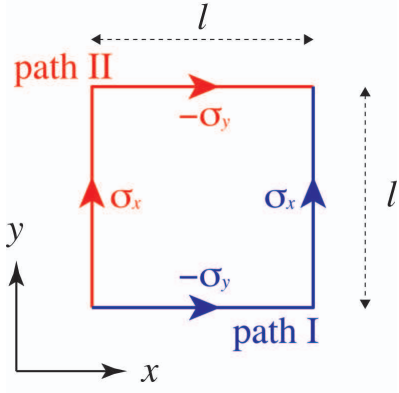}
\caption{An interference circuit on an infinitesimal $l\times l$ square.
The gauge fields are constant but non-commutative, which generates a phase difference between the two paths.}
\label{fig1}
\end{figure}

A simple way of finding a consequence of the non-Abelian gauge field is given as follows.
Consider the paths I and II on an infinitesimal $l\times l$ square as in Fig.~\ref{fig1}.
The phase factor which the electron taking the path I acquires is given by
\begin{eqnarray}
\label{eq80}
U_\mathrm{I}&\simeq&
e^{i\theta l\sigma_x/2}e^{-i\theta l\sigma_y/2}
\nonumber\\
&\simeq& e^{i\theta l(\sigma_x-\sigma_y)/2}e^{\theta^2 l^2 [\sigma_x,\sigma_y]/8},
\end{eqnarray}
while the phase factor on the path II is given by
\begin{eqnarray}
\label{eq90}
U_\mathrm{II}&\simeq&
e^{-i\theta l\sigma_y/2}e^{i\theta l\sigma_x/2}
\nonumber\\
&\simeq&
e^{i\theta l(\sigma_x-\sigma_y)/2}e^{-\theta^2 l^2 [\sigma_x,\sigma_y]/8},
\end{eqnarray}
where we used the lowest-order Baker-Campbell-Hausdorff formula for infinitesimal $l$.
Thus the phase difference between the paths, or equivalently the phase factor which the electron acquires upon circling the infinitesimal square is
\begin{equation}
\label{eq100}
U_\mathrm{phase}={U_\mathrm{II}}^\dag U_\mathrm{I}
\simeq e^{\theta^2 l^2[\sigma_x,\sigma_y]/4}
=e^{i\theta^2 l^2 \sigma_z/2}.
\end{equation}
This phase factor is caused by the non-commutativity of $\sigma_x$ and $\sigma_y$.
Interpreting that a flux $\Phi_R$ of a physical field generates the phase factor~(\ref{eq100}) in the form
\begin{equation}
\label{eq110}
U_\mathrm{phase}\simeq e^{i\theta\Phi_R/\hbar},
\end{equation}
we find that the physical field $\Phi_R/l^2$ is given by 
\begin{equation}
\label{eq115}
\frac{\Phi_R}{l^2}=\theta\frac{\hbar}{2}\sigma_z
\end{equation}
for infinitesimal $l$.
We thus realize that the non-Abelian gauge field actually generates a physical field.
Note here that the physical field is spin-dependent.

The above argument holds quantitatively only for the lowest order in $l$;
the Baker-Campbell-Hausdorff formula gives in Eq.~(\ref{eq80}) and~(\ref{eq90}) higher order terms containing $\sigma_x$ and $\sigma_y$.
This fact is important in realizing a perfect spin filter in Sec.~III below.

\subsection{Spin-orbit interaction and a vector potential}

Let us now generalize the above argument to the case where we have both a spin-orbit interaction and a vector potential.
The Hamiltonian with the vector potential and the Rashba interaction is given by~\cite{Engel06}
\begin{eqnarray}
\label{eq10}
\mathcal{H}_\mathrm{RSO}&=&\frac{1}{2m^\ast}\left({\Pi_x}^2+{\Pi_y}^2\right)
+\alpha\left(\Pi_x\sigma_y-\Pi_y\sigma_x\right)
\nonumber\\
\label{eq11}
&=&
\frac{1}{2m^\ast}
\left[
\left(p_x-eA_x(\vec{x})+\theta\frac{\hbar}{2}\sigma_y\right)^2
\right.
\nonumber\\
&&\phantom{\frac{1}{2m^\ast}}\left.
+\left(p_y-eA_y(\vec{x})-\theta\frac{\hbar}{2}\sigma_x\right)^2
\right]
+\mathrm{const.},
\nonumber\\
&&
\end{eqnarray}
where 
\begin{equation}
\vec{\Pi}\equiv \vec{p}-e\vec{A}(\vec{x})
\end{equation}
is the generalized momentum operator with the vector potential $\vec{A}(\vec{x})$.
Let us compare the Hamiltonian~(\ref{eq11}) with the Yang-Mills Hamiltonian~\cite{Ryder}
\begin{equation}
\label{eq12}
\mathcal{H}_\mathrm{YM}
=\frac{1}{2m}
\left[\left(p_x-\tilde{e}\tilde{A}_x(\vec{x})\right)^2
+\left(p_y-\tilde{e}\tilde{A}_y(\vec{x})\right)^2\right],
\end{equation}
where $\tilde{e}$ is the coupling constant and each of the gauge field components $\tilde{A}_\mu$ is actually a $2\times 2$ Hermitian matrix.
We can make the identification
\begin{eqnarray}
\label{eq41}
\tilde{A}_x(\vec{x}) &\leftrightarrow& eA_x(\vec{x})-\theta\frac{\hbar}{2}\sigma_y,
\quad\mbox{and}
\nonumber\\
\label{eq42}
\tilde{A}_y(\vec{x}) &\leftrightarrow& eA_y(\vec{x})+\theta\frac{\hbar}{2}\sigma_x.
\end{eqnarray}
We here set $\tilde{e}=1$ for later convenience.

The standard gauge field theory asserts that the physical field generated by the gauge field $\vec{\tilde{A}}$ is~\cite{Ryder}
\begin{eqnarray}
\label{eq49}
F_{\mu\nu}&=&i\frac{\hbar}{\tilde{e}}\left[D_\mu,D_\nu\right]
\nonumber\\
\label{eq50}
&=&\left(\frac{\partial}{\partial x_\mu}\tilde{A}_\nu-\frac{\partial}{\partial x_\nu}\tilde{A}_\mu\right)
-i\frac{\tilde{e}}{\hbar}\left[\tilde{A}_\mu,\tilde{A}_\nu\right],
\end{eqnarray}
where $D_\mu$ is the covariant derivative defined by
\begin{equation}
\label{eq60}
D_\mu=\frac{\partial}{\partial x_\mu}-i\frac{\tilde{e}}{\hbar}\tilde{A}_\mu(\vec{x})
\end{equation}
with $\mu,\nu=1,2$, $x_1\equiv x$ and $x_2\equiv y$.
The first term of the right-hand side of Eq.~(\ref{eq50}) is the contribution due to the rotation of the gauge field, while the last term of Eq.~(\ref{eq50}) is the contribution due to the non-Abelian nature of the Yang-Mills field.

Applying the formula~(\ref{eq50}) to Eq.~(\ref{eq41}), we find the physical field to be
\begin{equation}
\label{eq70}
F_{xy}=-F_{yx}
=eB_z+\theta^2\frac{\hbar}{2}\sigma_z
=e\frac{\Phi_B}{S}+\theta\frac{\Phi_R}{S},
\end{equation}
where $\Phi_B=SB_z$ is the usual magnetic flux penetrating the area $S$, while $\Phi_R$ is the flux of the physical field given by Eq.~(\ref{eq115}).
A parallel discussion shows that the Dresselhaus interaction gives the physical field
\begin{equation}
\label{eq75}
F_{xy}=eB_z-\theta^2\frac{\hbar}{2}\sigma_z
=e\frac{\Phi_B}{S}+\theta\frac{\Phi_D}{S},
\end{equation}
where $\Phi_D$ is the flux of the physical field generated by the Dresselhaus interaction.

In both Eqs.~(\ref{eq70}) and~(\ref{eq75}), the first term on the right-hand side is the usual magnetic field coming from a rotating vector potential, which causes the Aharonov-Bohm phase, and $e$ is the electron charge.
The second term on the right-hand side is the physical field generated by the non-commutativity of the Yang-Mills field, which causes the Aharonov-Casher phase, and $\theta$ plays the role of the charge for the spin-orbit interaction.
The two phases are often explained in the following context; a charged particle circulating a magnetic flux acquires the Aharonov-Bohm phase, whereas a magnetic moment circulating an electric flux acquires the Aharonov-Casher phase.
We emphasize that the standard non-Abelian gauge field theory explains the phases straightforwardly in the unified way shown in Eqs.~(\ref{eq70}) and~(\ref{eq75}).
It is our opinion that the explanation here is more general than the prevailing explanation of the Aharonov-Casher phase as a phase of a spin circulating an electric flux.
The Rashba coupling $\alpha$ is indeed proportional to the external electric field, but the Dresselhaus interaction is induced by a crystal field, not an external electric field.

We also note that the physical fields in Eqs.~(\ref{eq70}) and~(\ref{eq75}) have no cross terms between the vector potential and the spin-orbit interaction;
the position-dependent Abelian part of the gauge field (the vector potential) solely gives the Aharonov-Bohm phase, while the position-independent non-Abelian part of the gauge field (the spin-orbit interaction) solely gives the Aharonov-Casher phase.

A position-dependent spin-orbit interaction would result in two modifications of the above theory.
First, the constant term of the right-hand side of Eq.~(\ref{eq11}) would be a static potential instead.
Second, the first term of the right-hand side of Eq.~(\ref{eq50}) would include a contribution from the spin-orbit interaction, and hence the second term of the right-hand side of Eq.~(\ref{eq70}) would be  position-dependent.
The first modification, in particular, would result in changes of the energy dependence of the conductance below and hence might hinder the realization of the perfect spin filter discussed in Sec.~III.

\section{Perfect Spin Filter}
We again concentrate on the Rashba spin-orbit interaction hereafter.
We stress that the second term of the physical field~(\ref{eq70}) is spin-dependent.
When an up-spin electron feels a positive field, the down-spin electron feels a negative field of the same magnitude.
In contrast, the usual vector potential that generates the usual magnetic field (the first term of the physical field~(\ref{eq70})) is spin-independent;
it creates a phase factor of the same sign for both up spins and down spins.

Now let us adjust the parameters so that, when electrons circle an interference circuit, the spin-orbit interaction may create the phase $\pm \pi/2$ for up and down spins, respectively, \textit{and} the usual vector potential may create the additional phase $\pi/2$ for both spins.
Then the up-spin electrons acquire the phase factor $e^{i\pi}=-1$ while the down-spin electrons acquire the phase factor $1$;
the interference is completely destructive for the up-spin electrons and completely constructive for the down-spin electrons.
The up-spin electrons would not come out of such an interference circuit.
This constitutes a perfect spin filter.

In the following, we demonstrate that the above idea is indeed realized in two systems with a slight modification.
A difference of the reality from the above naive argument is due to the fact that Eq.~(\ref{eq100}) holds only in the lowest order of infinitesimal $l$.
As mentioned at the end of Sec.~IIA, other terms proportional to $\sigma_x$ and $\sigma_y$ appear in the phase for a circuit of finite size.
Therefore, the perfect spin filter is realized only for a tilted quantization axis which diagonalizes the phase factor $U_\mathrm{phase}$.
The idea of the perfect spin filter was presented in Refs.~\cite{Nitta99,Molnar04}, but both works, employing perturbation theory, did not elaborate the idea to the extent of tilting the spin quantization axis.
The tilt of the spin quantization axis was taken in Ref.~\cite{Hentschel04} to analyze a spin switch.

\subsection{An interference circuit of a tight-binding model}
\begin{figure}
\centering
\includegraphics[width=0.95\columnwidth]{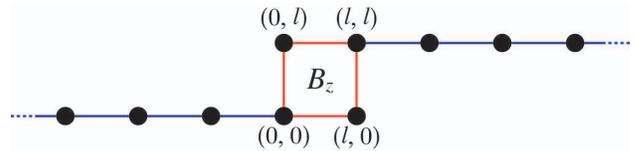}
\caption{A tight-binding model on the square has the Rashba interaction.
A magnetic flux is penetrating the square.
Two leads of tight-binding chains free of the Rashba interaction are attached to the left of the point $(0,0)$ and to the right of the point $(l,l)$.}
\label{fig2}
\end{figure}

As a first example, we consider a tight binding model on a lattice shown in Fig.~\ref{fig2}.
The four edges on the central square are subject to the Rashba spin-orbit interaction, while the left lead ($x\leq0$) and the right lead ($x\geq l$) are free of the Rashba interaction.
The magnetic flux $B_z$ penetrating the square can be described by the vector potential
\begin{equation}
\label{eq120}
\vec{A}=\left(-\frac{1}{2}B_z y, \frac{1}{2}B_z x,0\right).
\end{equation}
Assuming that the magnetic field is contained inside the square and not applied on the paths themselves, we neglect the Zeeman term here.

We use the standard definition of the tight-binding model with the Rashba interaction~\cite{Ando92} for the square.
The hopping from $(0,0)$ to $(l,0)$ generates the phase factor $e^{-i\theta l\sigma_y/2}$.
Hence we have the hopping term
\begin{equation}
\label{eq130}
-tc^\dag_{(l,0)}e^{-i\theta l\sigma_y/2}c_{(0,0)}
=-tc^\dag_{(l,0)}\left(c_\theta-i s_\theta \sigma_y\right)c_{(0,0)},
\end{equation}
where 
\begin{eqnarray}
c_\theta&\equiv&\cos\frac{\theta l}{2},
\\
s_\theta&\equiv&\sin\frac{\theta l}{2}
\end{eqnarray}
and $c^\dag$ and $c$ are the creation and annihilation operators that have two components:
\begin{equation}
\label{eq140}
c^\dag_{(x,y)}\equiv
\left(\begin{array}{cc}
c^\dag_{(x,y)\uparrow} & c^\dag_{(x,y)\downarrow}
\end{array}\right),
\quad
c_{(x,y)}\equiv
\left(\begin{array}{c}
c_{(x,y)\uparrow} \\
c_{(x,y)\downarrow}
\end{array}\right).
\end{equation}
The hopping from $(0,0)$ to $(0,l)$ is described by the term
\begin{equation}
\label{eq150}
-tc^\dag_{(0,l)}e^{i\theta l\sigma_x/2}c_{(0,0)}
=-tc^\dag_{(0,l)}\left(c_\theta+i s_\theta \sigma_x\right)c_{(0,0)}.
\end{equation}
The hopping from $(0,l)$ to $(l,l)$ has, in addition to the above hopping elements, a phase due to the vector potential~(\ref{eq120}) of the form $e^{-iel^2B_z/(2\hbar)}$ and the hopping from $(l,0)$ to $(l,l)$ has the additional phase $e^{iel^2B_z/(2\hbar)}$.
The leads attached to the points $(0,0)$ and $(l,l)$, on the other hand, are given by the simple tight-binding model:
\begin{eqnarray}
\label{eq222}
&&-t\sum_{n=-\infty}^0
\left(
c^\dag_{(nl,0)}c_{((n-1)l,0)}
+c^\dag_{((n-1)l,0)}c_{(nl,0)}
\right)
\nonumber\\
&&-t\sum_{n=1}^\infty
\left(
c^\dag_{((n+1)l,0)}c_{(nl,0)}
+c^\dag_{(nl,0)}c_{((n+1)l,0)}
\right).
\end{eqnarray}

\begin{figure}
\centering
\includegraphics[width=0.9\columnwidth]{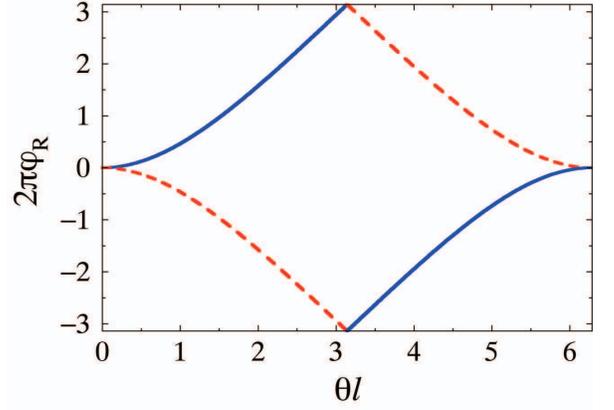}
\caption{The phases of the ``up'' spin (the solid line) and of the ``down'' spin (the broken line) due to the Rashba interaction of the tight-binding model on a square.}
\label{fig25}
\end{figure}

The above definitions of the hopping elements give the phase factor that
the spin-up and spin-down electrons at $(0,0)$,
\begin{equation}
\label{eq155}
\left(\begin{array}{c}
c_{(0,0)\uparrow}^\dag \\
c_{(0,0)\downarrow}^\dag
\end{array}\right)
\left|\mbox{vac.}\right\rangle,
\end{equation}
acquire upon circling the square, in the form
\begin{eqnarray}
\label{eq160}
U_\mathrm{phase}
&=&e^{2\pi i\varphi_B}(c_\theta-is_\theta\sigma_x)(c_\theta+is_\theta\sigma_y)
\nonumber\\
&&\times(c_\theta+is_\theta\sigma_x)(c_\theta-is_\theta\sigma_y),
\end{eqnarray}
where
\begin{equation}
\label{eq170}
\varphi_B\equiv \frac{el^2 B_z}{h} =\frac{\Phi_B}{\Phi_0}
\end{equation}
with $\Phi_B\equiv l^2 B_z$ being the magnetic flux and $\Phi_0$ the flux unit.
The eigenvalues of the $2\times 2$ matrix of the phase factor~(\ref{eq160}) is given by
\begin{equation}
\label{eq180}
e^{2\pi i(\varphi_B\pm\varphi_\mathrm{R})},
\end{equation}
where
\begin{equation}
\label{eq190}
\varphi_\mathrm{R}\equiv \frac{1}{2\pi}\arccos\left(1-2\sin^4\frac{\theta l}{2}\right).
\end{equation}
The phase $\varphi_\mathrm{R}$ due to the Rashba interaction is consistent with Eq.~(\ref{eq100}) for small $l$;
see Fig.~\ref{fig25}.

Now, adjust the parameters so that
\begin{equation}
\label{eq200}
2\pi\varphi_B=2\pi\varphi_\mathrm{R}=\frac{\pi}{2}
\end{equation}
may hold, or
\begin{eqnarray}
\label{eq210}
l^2B_z&=&\frac{\Phi_0}{4}
\quad
\mbox{and}
\\
\label{eq211}
\theta l&=&2 \arcsin \frac{1}{2^{1/4}}=1.997\cdots.
\end{eqnarray}
Then the phase factor is completely destructive for ``up'' spins and completely constructive for ``down'' spins, where the ``up'' spin and the ``down'' spin are defined as the ones that diagonalize $U_\mathrm{phase}$.
In the particular case~(\ref{eq200}), the ``up'' spin and the ``down'' spin are given by
\begin{eqnarray}
\label{eq220}
\tilde{\chi}_+&=&
\left(\begin{array}{c}
2^{-1/4}e^{-i\pi/4} \\
-\sqrt{1-1/\sqrt{2}}
\end{array}\right)
\nonumber\\
&=&\left(\begin{array}{c}
(1-i)  \times 0.595 \cdots\\
-0.541 \cdots
\end{array}\right)
\quad\mbox{and}
\\
\label{eq221}
\tilde{\chi}_-&=&
\left(\begin{array}{c}
\sqrt{1-1/\sqrt{2}} \\
2^{-1/4}e^{i\pi/4}
\end{array}\right)
\nonumber\\
&=&\left(\begin{array}{c}
0.541\cdots \\
(1+i) \times 0.595 \cdots
\end{array}\right),
\end{eqnarray}
respectively.
We note that the spin quantization axis of these bases is considerably tilted.

We then numerically calculated the transmission coefficients from the left lead to the right lead.
We employed the method using the self-energy of the leads~\cite{Datta,Sasada05}.
The transmission coefficient between a spin state $\lambda$ of the left lead and a spin state $\tau$ of the right lead at the energy $E$ is given by
\begin{equation}
\label{eq224}
T_{\tau\lambda}(E)=\left|
\left\langle\mathrm{vac.}\right|
c_{(l,l)\tau}
\frac{\sqrt{4t^2-E^2}}{E-\mathcal{H}_\mathrm{sq}-\Sigma(E)}
c^\dag_{(0,0)\lambda}
\left|\mathrm{vac.}\right\rangle
\right|,
\end{equation}
where $\mathcal{H}_\mathrm{sq}$ denotes the Hamiltonian of the central square and $\Sigma(E)$ denotes the self-energy defined by
\begin{equation}
\label{eq225}
\Sigma(E)\equiv\frac{E-i\sqrt{4t^2-E^2}}{2}
\left(c^\dag_{(0,0)}c_{(0,0)}+c^\dag_{(l,l)}c_{(l,l)}\right).
\end{equation}
The formula~(\ref{eq224}) requires inversion of only an $8\times8$ matrix.

\begin{figure}
\centering
\includegraphics[width=0.9\columnwidth]{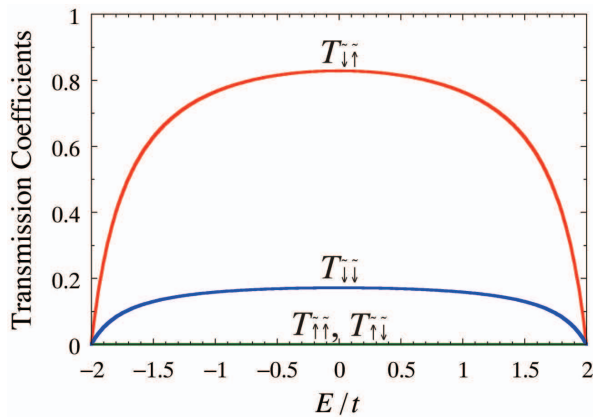}
\caption{The transmission coefficients from the left lead to the right lead of Fig.~\ref{fig2} in the case~(\ref{eq200}).
The conductances with the output of ``up'' spins are zero over the entire energy range.}
\label{fig3}
\end{figure}

Figure~\ref{fig3} shows the transmission coefficients in the case~(\ref{eq200}).
Here $T_{\tilde{\uparrow}\tilde{\downarrow}}$, for example, denotes the transmission coefficient of the ``down'' spin on the left lead and the ``up'' spin on the right lead.
(Remember that the ``up'' spin and the ``down'' spin are defined by Eqs.~(\ref{eq220}) and~(\ref{eq221}), respectively.)
Under the condition~(\ref{eq200}), the ``up'' spin electrons indeed do not come out of the circuit to the right lead for any energy.
Thus we realized a perfect spin filter.

\begin{figure}
\centering
\includegraphics[width=0.95\columnwidth]{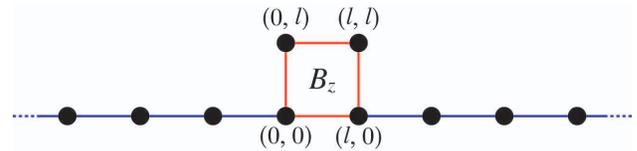}
\caption{An asymmetric circuit, where the right lead is attached to the point $(l,0)$ instead of the point $(l,l)$ as in the symmetric circuit in Fig.~\ref{fig2}.}
\label{fig31}
\end{figure}

\begin{figure}
\centering
\includegraphics[width=0.9\columnwidth]{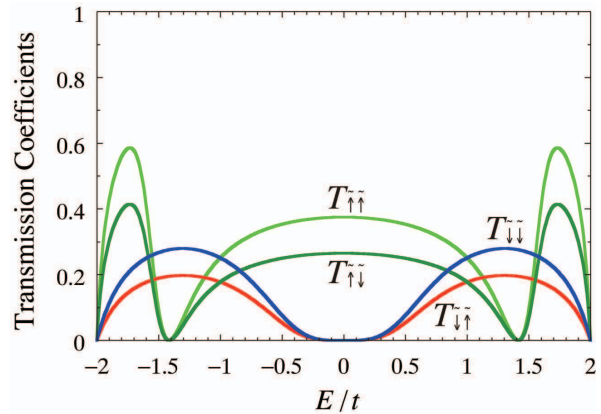}
\caption{The transmission coefficients from the left lead to the right lead of Fig.~\ref{fig31} in the case~(\ref{eq200}).
The conductances with the output of ``up'' spins are zero only at particular points of the energy.}
\label{fig32}
\end{figure}

The geometric symmetry between the lower and upper arms of the interference circuit in Fig.~\ref{fig2} does not seem to play an apparent role in the above argument.
Inspection of details of the calculation would, in fact, reveal that the symmetry is crucial in realizing the perfect spin filter for any energy.
Let us consider, for example, the asymmetric circuit shown in Fig.~\ref{fig31} for the same values of the parameter as in Eqs.~(\ref{eq210})--(\ref{eq221}).
The result in Fig.~\ref{fig32} shows that the perfect spin filter is realized only at particular points of the energy.

\subsection{A quantum ring}
\begin{figure}
\centering
\includegraphics[width=0.8\columnwidth]{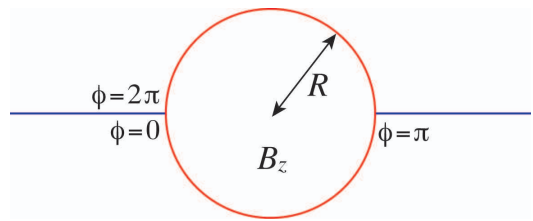}
\caption{A quantum ring of radius $R$ with the spin-orbit interaction with two leads of quantum wires without the spin-orbit interaction attached to the points $\phi=0$ and $\phi=\pi$.
A magnetic field $B_z$ penetrates the ring.}
\label{fig45}
\end{figure}

As a second example of the perfect spin filter, we consider a quantum ring of radius $R$ with two quantum wires attached (Fig.~\ref{fig45}).
The Hamiltonian of the quantum ring ($0\leq\phi\leq 2\pi$) with the Rashba spin-orbit interaction is given by~\cite{Zhou94,Meijer02}
\begin{equation}
\label{eq230}
\mathcal{H}_\mathrm{ring}=\frac{\hbar^2}{2m^\ast R^2}
\left(-i\frac{\partial}{\partial\phi}-\varphi_B-\frac{\theta R}{2}\sigma_r\right)^2,
\end{equation}
where
\begin{eqnarray}
\label{eq240}
\varphi_B&\equiv& \frac{e\pi R^2B_z}{h}=\frac{\Phi_B}{\Phi_0},
\\
\label{eq241}
\theta&\equiv&\frac{2m^\ast\alpha}{\hbar},
\quad\mbox{and}
\\
\label{eq242}
\sigma_r&\equiv&\sigma_x\cos\phi+\sigma_y\sin\phi.
\end{eqnarray}
The leads attached to the quantum ring at $\phi=0$ and $\phi=\pi$ are simple quantum wires free of the Rashba interaction:
\begin{equation}
\label{eq243}
\mathcal{H}_\mathrm{leads}=-\frac{\hbar^2}{2m^\ast}
\frac{\partial^2}{\partial x^2}.
\end{equation}

\begin{figure}
\centering
\includegraphics[width=0.9\columnwidth]{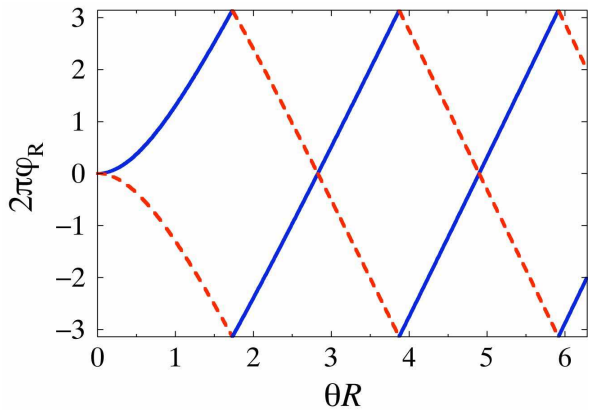}
\caption{The phases of the ``up'' spin (the solid line) and of the ``down'' spin (the broken line) due to the Rashba interaction of a quantum ring.}
\label{fig35}
\end{figure}

We can exactly obtain the eigenfunction of the ring Hamiltonian~(\ref{eq230}) in the form~\cite{Yi97}
\begin{equation}
\label{eq250}
\Psi_{\pm\pm}(\phi;k_\phi)
=e^{i(\pm k_\phi+\varphi_B\pm \varphi_\mathrm{R})\phi}e^{-i\beta\sigma_\phi/2}\chi_\pm
\end{equation}
with the eigenvalues
\begin{equation}
\label{eq255}
E=\frac{\hbar^2{k_\phi}^2}{2m^\ast R^2},
\end{equation}
where
\begin{eqnarray}
\label{eq260}
\varphi_\mathrm{R}&\equiv&\frac{1}{2}\left(\sqrt{1+\theta^2R^2}-1\right),
\\
\label{eq261}
\sigma_\phi&\equiv&\sigma_y\cos\phi-\sigma_x\sin\phi,
\\
\label{eq262}
\beta&\equiv& \arctan\theta R
\end{eqnarray}
and $\chi_\pm$ are the eigenfunctions of $\sigma_z$.
The first sign of the subscript of $\Psi_{\pm\pm}$ denotes the sign of the momentum and the second sign denotes the sign of the spin.
The phase $\varphi_\mathrm{R}$ due to the Rashba interaction is consistent with Eq.~(\ref{eq100}) for small $R$;
see Fig.~\ref{fig35}.
The wave function~(\ref{eq250}) at $\phi=2\pi$ is
\begin{equation}
\label{eq270}
\Psi_{\pm\pm}(2\pi;k_\phi)=e^{\pm 2\pi i k_\phi}U_\mathrm{phase}\chi_\pm
\end{equation}
with the phase factor
\begin{equation}
\label{eq280}
U_\mathrm{phase}\equiv
e^{2\pi i (\varphi_B\pm\varphi_\mathrm{R})}e^{-i\beta\sigma_y/2}.
\end{equation}

We now realize the perfect spin filter by adjusting the parameters so that
\begin{equation}
\label{eq290}
2\pi\varphi_B=\frac{\pi}{2}+2m\pi
\quad\mbox{and}\quad
2\pi\varphi_\mathrm{R}=\frac{\pi}{2}+2n\pi
\end{equation}
may hold, where $m$ and $n$ are integers.
The condition~(\ref{eq290}) translates to
\begin{equation}
\label{eq300}
\pi R^2 B_z \equiv \frac{1}{4}\Phi_0,\frac{5}{4}\Phi_0,\frac{9}{4}\Phi_0,\ldots
\end{equation}
for $m=0,1,2,\ldots$ and 
\begin{equation}
\label{eq310}
\theta R\equiv\frac{\sqrt{5}}{2},\frac{\sqrt{45}}{2},\frac{\sqrt{117}}{2},\ldots
\end{equation}
for $n=0,1,2,\ldots$.
The eigenfunctions of the phase factor $U_\mathrm{phase}$ are
\begin{equation}
\tilde{\chi}_\pm= e^{-i\beta\sigma_y/2}\chi_\pm
\end{equation}
and, in the case~(\ref{eq290}), are
\begin{eqnarray}
\label{eq320}
\tilde{\chi}_+&=&\frac{1}{\sqrt{6}}
\left(\begin{array}{c}
\sqrt{5} \\
1
\end{array}\right)
=\left(\begin{array}{c}
0.913\cdots \\
0.408\cdots
\end{array}\right)
\quad\mbox{and}
\\
\label{eq321}
\tilde{\chi}_-&=&\frac{1}{\sqrt{6}}
\left(\begin{array}{c}
-1 \\
\sqrt{5}
\end{array}\right)
=\left(\begin{array}{c}
-0.408\cdots \\
0.913\cdots
\end{array}\right).
\end{eqnarray}
Again, the spin quantization axis of these bases is considerably tilted.

We calculated the transmission coefficients following Ref.~\cite{Yi97}.
We first assume the amplitudes of the left-going and right-going wave functions separately for the left lead, the portion $0<\phi<\pi$ of the ring, the portion $\pi<\phi<2\pi$ of the ring, and the right lead.
This amounts to sixteen amplitudes in total when we take the spin degree of freedom into account.
The continuation of the wave function at $\phi=0$ and $\phi=\pi$ give eight conditions and the conservation of the generalized momentum at $\phi=0$ and $\phi=\pi$ give four conditions.
We thus end up with four degrees of freedom.
We obtain the $S$ matrix of the quantum ring by expressing the four amplitudes of the out-going waves (the left-going wave on the left lead and the right-going wave on the right lead with spin up and down) in terms of the four amplitudes of the in-coming waves (the right-going wave on the left lead and the left-going wave on the right lead with spin up and down).
The off-diagonal $2\times2$ block of the $4\times4$ $S$ matrix give the transmission coefficients.

\begin{figure}
\centering
\includegraphics[width=0.9\columnwidth]{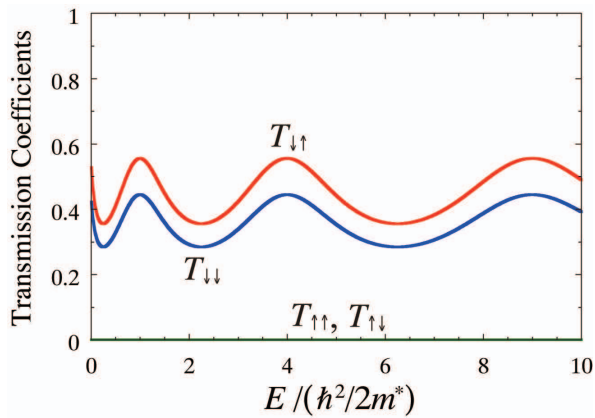}
\caption{The transmission coefficients from the left lead to the right lead in the case~(\ref{eq290}).
The conductances with the output of ``up'' spins are identically zero over the entire energy range.}
\label{fig4}
\end{figure}

Algebra shows us that the transmission coefficients have the factors
\begin{eqnarray}
\label{eq330}
T_{\tilde{\uparrow}\tilde{\uparrow}},T_{\tilde{\uparrow}\tilde{\downarrow}}
&\propto&\left|1+e^{2\pi i(\varphi_B+\varphi_\mathrm{R})}\right|^2
\quad\mbox{and}
\\
\label{eq340}
T_{\tilde{\downarrow}\tilde{\uparrow}},T_{\tilde{\downarrow}\tilde{\downarrow}}
&\propto&\left|1+e^{2\pi i(\varphi_B-\varphi_\mathrm{R})}\right|^2,
\end{eqnarray}
where $\tilde{\uparrow}$ and $\tilde{\downarrow}$ denote, respectively, the ``up'' spin $\tilde{\chi}_+$ and ``down'' spin $\tilde{\chi}_-$ diagonalizing the phase factor $U_\mathrm{phase}$.
Hence we \textit{identically} have $T_{\tilde{\uparrow}\tilde{\uparrow}}=T_{\tilde{\uparrow}\tilde{\downarrow}}=0$ in the case~(\ref{eq290}), which is demonstrated in Fig.~\ref{fig4}.
We again achieved a perfect spin filter.

Let us estimate the parameters for the condition~(\ref{eq290}).
Using the experimental values
\begin{eqnarray}
\label{eq350}
\alpha\hbar&=&3\times10^{-11}\mbox{[eV m]}
\quad
\mbox{and}
\\
m^\ast&=&0.041m_e
\end{eqnarray}
for an InGaAs/InAlAs heterojunction~\cite{Sato01}, we have
\begin{eqnarray}
\label{eq360}
R&\simeq& 40 \mbox{[nm]}, 110 \mbox{[nm]}, 180 \mbox{[nm]},\ldots
\quad\mbox{and}
\\
\label{eq361}
B_z&\simeq& 0.15\mbox{[T]}, 0.02\mbox{[T]}, 0.008\mbox{[T]},\ldots
\end{eqnarray}
for $m=0$ and $n=0,1,2,\ldots$.
The ring radius is quite less than the one in a recent experiment~\cite{Konig06} observing the Aharonov-Casher phase, but may be achievable in experiments.

\section{Summary}
In the present article, we showed that the Rashba spin-orbit interaction can be analyzed with a non-Abelian gauge field theory.
The Aharonov-Bohm phase and the Aharonov-Casher phase are discussed in a unified way.
The analysis gives an idea for realizing a perfect spin filter.
We indeed achieved perfect spin filters in two interference circuits.

After submitting the first version of the present manuscript, we noticed a recent paper on a spin filter~\cite{Citro06}.
They considered the same situation of the quantum ring but realized a spin filter only at separate points of the energy; see Fig.~2 of Ref.~\cite{Citro06}.
We presume that this is because they did not tilt the quantization axis of the spin.

\section*{Acknowledgments}
The work is supported partly by Grand-in Aid for Exploratory Research (No.~17654073) from the Ministry of Education, Culture, Sports, Science and Technology and partly by the Murata Science Foundation as well as by the National Institutes of Natural Sciences undertaking Forming Bases for Interdisciplinary and International Research through Cooperation Across Fields of Study and Collaborative Research Program (No. NIFS06KDBT005).
One of the authors (N.H.) acknowledges support by Grant-in-Aid for Scientific Research (No.~17340115) from the Ministry of Education, Culture, Sports, Science and Technology as well as support by Core Research for Evolutional Science and Technology of Japan Science and Technology Agency.

\end{document}